\documentclass{aa}  
\pdfoutput=1

\usepackage{amsmath}
\usepackage{amssymb}
\usepackage{epsfig}
\usepackage{multirow}
\usepackage{longtable}
\usepackage{lscape}
\usepackage{bbm}
\usepackage{dcolumn}
\usepackage{rotating}
\usepackage[squaren]{SIunits}
\usepackage{graphicx}
\usepackage{txfonts}
\newcommand{\bi}{\begin{itemize}}
\newcommand{\ei}{\end{itemize}}

\newcommand{\be}{\begin{equation}}
\newcommand{\ee}{\end{equation}}
\newcommand{\bea}{\begin{eqnarray}}
\newcommand{\eea}{\end{eqnarray}}

\newcommand{\ie}{{\it i.e.}}

\newcommand{\eg}{{\it e.g.}}

\newcommand{\cf}{{\it cf.}}

\newcommand{\eq}{Eq.}

\newcommand{\fig}{Fig.}

\newcommand{\Tab}{Table}

\newcommand{\equ}[1]{\eq~(\ref{equ:#1})}
\newcommand{\figu}[1]{\fig~\ref{fig:#1}}

\newcommand{\capdef}{}
\newcommand{\mycaption}[2][\capdef]{\renewcommand{\capdef}{#2}%
        \caption[#1]{{\footnotesize #2}}}
\makeatletter
\renewcommand{\fnum@table}{\textbf{\tablename~\thetable}}
\renewcommand{\fnum@figure}{\textbf{\figurename~\thefigure}}
\makeatother

\begin{document} 
   \title{Impact of Secondary Acceleration in Gamma-Ray Bursts}

%   \subtitle{}

   \author{W.\ Winter
          \inst{1}
          \and
         J.\ Becker Tjus\inst{2}
\and
S.\ R.\ Klein\inst{3,4}}

   \institute{DESY, Platanenallee 6, 15738 Zeuthen, Germany\\
              \email{walter.winter@desy.de}
         \and
Theoretische Physik IV: Plasma-Astroteilchenphysik, Fakult\"at f\"ur Physik \& Astronomie, Ruhr-Universit\"at Bochum 44780 Bochum, Germany\\
             \email{julia.tjus@rub.de}
\and
Lawrence Berkeley National Laboratory, Berkeley CA, 94720, USA \\
\email{srklein@lbl.gov}
\and
Department of Physics, University of California, Berkeley, CA, 94720 USA\\}
 
  \date{}

% \abstract{}{}{}{}{} 
% 5 {} token are mandatory
 
  \abstract
   {It is clear from the observation of charged cosmic rays up to energies of $10^{20}$~eV that particle acceleration must occur in astrophysical sources. Acceleration of secondary particles like muons and pions, produced in cosmic ray interactions, are usually neglected, however, when calculating the flux of neutrinos from cosmic ray interactions.}
   {
Here, we discuss the acceleration of secondary muons, pions, and kaons in gamma-ray bursts within the internal shock scenario, and their impact on the neutrino fluxes.}
   {We introduce a two-zone model consisting of an acceleration zone (the shocks) and a radiation zone (the plasma downstream the shocks). 
The acceleration in the shocks, which is an unavoidable consequence of the efficient proton acceleration,  requires efficient transport from the radiation back to the acceleration zone. On the other hand, stochastic acceleration in the radiation zone can enhance the secondary spectra of muons and kaons significantly if there is a sufficiently large turbulent region.}
   {Overall, it is plausible that neutrino spectra can be enhanced by up to a factor of two at the peak by stochastic acceleration, that an additional spectral peaks appears from shock acceleration of the secondary muons and pions, and that the neutrino production from kaon decays is enhanced.  }
   {Depending on the GRB parameters, the general conclusions concerning the limits to the internal shock scenario obtained by recent IceCube and ANTARES analyses may be affected by up to a factor of two by secondary acceleration.
Most of the changes occur at energies above $10^7$~GeV, so the effects for next-generation radio-detection experiments will be more pronounced.
In the future, however, if GRBs are detected as high-energy neutrino sources, the detection of one or several pronounced peaks around $10^{6}$~GeV or higher energies could help to derive the basic properties of the magnetic field strength in the GRB.}

   \keywords{Acceleration of particles --
                Neutrinos --
                Astroparticle Physics --
		Gamma-ray burst: general
               }

   \maketitle
%
%________________________________________________________________

\section{Introduction}

Gamma-ray bursts (GRBs) are 
a candidate class for the origin of the ultra-high energy cosmic rays (UHECRs). A popular scenario is the internal shock model, where the prompt $\gamma$-ray emission originates from the radiation of particles accelerated by internal shocks in the ejected material~
\citep{Paczynski:1994uv,Rees:1994nw} (see \cite{Piran:2004ba,Meszaros:2006rc} for reviews). 
If a significant 
baryon flux  is accelerated, the GRBs may be a plausible source for the UHECRs. In this case, substantial production of secondary pions, muons, and also kaons 
are expected from photohadronic interactions between the baryons and the radiation field; these will decay into neutrinos and other decay products~\citep{Waxman:1997ti,Asano:2006zzb}. 
 Gamma-rays at TeV energies and above are co-produced in the photohadronic process, but are subject  to interactions with the internal photon field from the radiation processes, including synchrotron radiation and inverse Compton scattering, in the GRB. The photons are expected to cascade down via pair production cascades so that they can be detected at $\sim$ GeV energies. Several of such GRBs have been detected be Fermi-LAT \citep{fermi_bursts}, but the associated neutrino production per burst is generally expected to be rather low, see e.g.\ \cite{bhoo2010}. In general, neutrino detection from GRBs with IceCube therefore needs to be done via the stacking of a larger number of bursts, see e.g.\ \cite{Abbasi:2011qc,Abbasi:2012zw}.

Very stringent neutrino flux limits for 
the internal shock scenario have been recently obtained by the IceCube collaboration using the stacking approach~\citep{Abbasi:2011qc,Abbasi:2012zw}.  Using timing, energy, and directional information for the individual bursts, new limits have been obtained, which are basically background-free and which are significantly below earlier predictions 
based on gamma-ray observations~\citep{Waxman:1997ti,Guetta:2003wi,Becker:2005ej,Abbasi:2009ig}. These predictions have been recently revised from the theoretical perspective~\citep{Hummer:2011ms,Li:2011ah,He:2012tq}, yielding about a factor of ten lower expected flux~\citep{Hummer:2011ms} depending on the analytical method compared to; see also \cite{Adrian-Martinez:2013sga} for an analysis by the ANTARES collaboration using this method. This discrepancy comes mainly from the energy dependence of the mean free path of the protons, the integration over the full photon target spectrum (instead of using the break energy for the pion production efficiency), and several other corrections adding up in the same direction; see Fig.~1 (left) in \cite{Hummer:2011ms}.
 It should be noted at this point, that the absolute normalization of the neutrino flux scales linearly with the ratio of the luminosity in protons to electrons (baryonic loading). In typical models \citep{Waxman:1997ti,Guetta:2003wi,Becker:2005ej,Abbasi:2009ig}, this ratio is usually assumed to be $10$, while theoretical considerations %SK rather 
suggest a value of $100$ \citep{schlickeiser2002} if GRBs are to be the sources of the UHECRs. In a recent study~\citep{Baerwald:2014zga}, this value is self-consistently derived from the combined UHECR source and propagation model, including the  fit of the UHECR data. For an injection index of two, it is demonstrated that this value depends on the burst parameters, and that values between $10$ and $100$ are plausible.\footnote{A baryonic loading of $10$ requires, however, ``typical'' source parameters $\Gamma \sim 400$ and $L_{\gamma,\mathrm{iso}} \sim 10^{53} \, \mathrm{erg \, s^{-1}}$, whereas $\Gamma \sim 300$ and $L_{\gamma,\mathrm{iso}} \sim 10^{52} \, \mathrm{erg \, s^{-1}}$ more point towards a baryonic loading of order $100$; see Fig.~7 in \cite{Baerwald:2014zga}.}
Note that these numbers depend strongly on the proton and electron/photon input spectral shapes and energy ranges, and according to basic theory of stochastic acceleration, it can easily vary between $1000$ and $0.1$~\citep{lukas_paper}. As further demonstrated in \cite{Baerwald:2014zga}, the improved modeling of the GRB spectra  can be used to constrain the central parameters of the calculation, \ie, the ratio of protons to electrons and the boost factor. The effects discussed in our study could then contribute to determining another basic property of the GRB, namely the magnetic field strength.

Another argument can be used when relating
the neutrino and UHECR fluxes directly if the cosmic rays escape as
neutrons produced in the same interactions as the
neutrinos~\citep{Ahlers:2011jj}. Since this possibility is strongly
disfavored~\citep{Abbasi:2012zw} 
it is conceivable that other escape
mechanisms dominate for UHECR escape from
GRBs~\citep{Baerwald:2013pu}. For instance, if the Larmor radius can
reach the shell width at the highest energies, it is plausible
that a fraction of the cosmic rays can directly escape. Other possible
mechanisms include diffusion out of the shells. In \cite{Baerwald:2014zga},
it was demonstrated that even current IceCube data already
imply that these alternative escape mechanisms must dominate if GRBs ought to be
the sources of the UHECR, and that future IceCube data will exert
pressure on these alternative options as well.

The secondary pions, muons, and kaons produced by photohadronic interactions will typically either decay (at low energies) or lose energy by synchrotron radiation (at high energies). At the point where decay and synchrotron timescales are equal, a spectral break in the secondary, and therefore also 
in the neutrino spectrum is expected.  This is the so-called ``cooling break'',
see e.g.\ \cite{Waxman:1997ti}. Additional processes, which potentially affect the secondary spectra, are: adiabatic losses, interactions with the radiation field~\citep{Kachelriess:2007tr}, and acceleration of the secondaries in the shocks or by stochastic acceleration~\citep{Koers:2007je,Murase:2011cx,Klein:2012ug}. In this study, we focus on the quantitative impact of the secondary acceleration on the neutrino fluxes, and the conditions for a significant contribution of this effect. 
A substantial enhancement of the secondary spectra would increase the tension between the recent IceCube observations and the predictions, and would therefore be critical for the interpretation of the recent IceCube results.

%%%%%%%%%%%%%%%%%%%%%%%%%%%%%%%%%%
\section{Model description}
%%%%%%%%%%%%%%%%%%%%%%%%%%%%%%%%%%

\begin{figure*}[t]
\begin{center}
\begin{tabular}{ccc}
\includegraphics[width=0.32\textwidth]{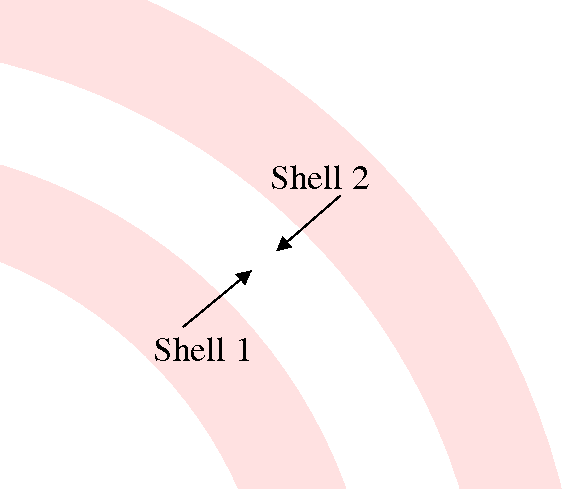} &
\includegraphics[width=0.32\textwidth]{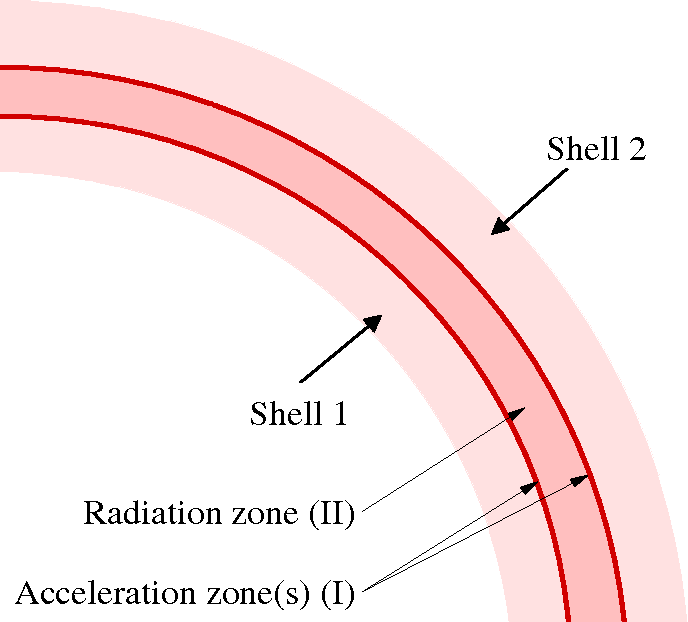} & 
\includegraphics[width=0.32\textwidth]{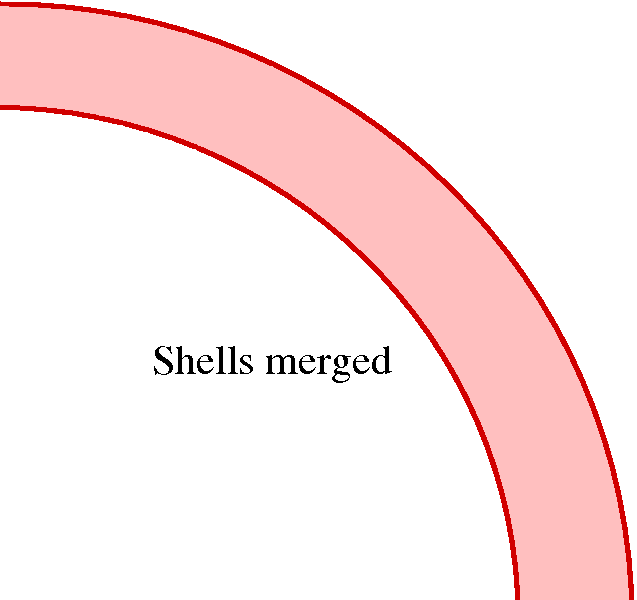} \\[0.5cm]
(i) & (ii) & (iii)
\end{tabular}
 \end{center}
\mycaption{\label{fig:shells} Collision of two shells, illustrated, assuming that they merge (irrelevant for the model presented here).
}
\end{figure*}

The effect of linear acceleration on the secondaries has been discussed in \cite{Klein:2012ug}, where it 
was demonstrated that significant acceleration effects can be expected if the secondaries can pile up over a large enough energy range. In GRBs, one has shock 
(Fermi 1st order) acceleration and, possibly, stochastic 
(Fermi 2nd order) acceleration in the plasma downstream the shock if turbulent magnetic fields are present; see \cite{Murase:2011cx}, where qualitative estimates for the secondary acceleration are made. An important effect is that the secondaries will be mostly produced by photohadronic interactions downstream
of
the shock, where high photon and proton densities are available over the dynamical timescale of the collision 
$t'_{\mathrm{dyn}}$ (as usual, we refer to quantities in the shock rest frame, SRF, by primed quantities). 
We therefore propose a two-zone model, including:
\begin{description}
\item[Acceleration zone (I):] Forward and reverse shocks.
\item[Radiation zone (II):] Plasma downstream the shocks.
\end{description}
The collision of the shells in the internal shock model is illustrated in \figu{shells}, where the different zones are shown as well. Although we do not consider the explicit time dependence, it is a good approximation to use a steady 
state model with constant effective densities over the dynamical timescale $t'_{\mathrm{dyn}}$.
This dynamical timescale is typically related to the width of the shells $\Delta r'$ by
\begin{equation} 
c t'_{\mathrm{dyn}}  \simeq \Delta r' \simeq \Gamma \, c \, \frac{t_v}{1+z}   \, ,
\label{equ:dr}
\end{equation}
where $t_v$ is the (observed) variability timescale
and $\Gamma$ is the appropriately averaged Lorentz boost of the shells, see \cite{Kobayashi:1997jk}. We henceforth assume that the adiabatic cooling timescale  is of the same order of magnitude $t'_{\mathrm{ad}} \sim t'_{\mathrm{dyn}}$.

The description of the secondary acceleration in GRBs faces several challenges. First of all, all species (pions, muons, and kaons) may be accelerated. Second, muons are produced by pion decays, which may be accelerated themselves. Third, it is expected that an efficient secondary acceleration is a consequence of the primary (proton, electron) acceleration. The amount of accelerated secondaries depends on the transport between radiation and acceleration zones. And fourth, the spectra of the secondaries, which originate from photohadronic interactions between protons and radiation, are no trivial power laws. It is therefore {\em a priori} not clear over what energy range the secondaries could pile up, and what the impact of spectral effects is. Our model aims to address these issues in a way as self-consistent as technically feasible, using state-of-the-art technology.

For the description of the target photon spectrum, we chose a framework relying on gamma-ray observations: it is assumed that the observed gamma-ray spectrum is representative for the spectrum within the source. 
Based on energy partition arguments, the baryonic and magnetic field densities can be obtained from the gamma-ray spectrum, and the secondary and neutrino fluxes can be computed from the photohadronic interactions between matter and radiation fields, see \cite{Guetta:2003wi,Becker:2005ej,Abbasi:2009ig}. 
This framework is slightly different from completely self-consistent (theoretical) approaches generating the target photon spectrum from the radiation processes such as synchrotron radiation and inverse Compton scattering of co-accelerated electrons, or photon production from $\pi^0$ decays, see \eg\ \cite{Asano:2011pn,Asano:2012jr}. The advantage of this approach is that it matches the gamma-ray observations by construction, but the drawback is that it cannot explain them. The overall parameters (photon density, $B'$) are assumed to be similar for both zones.

We include additional pion and kaon production modes using the methods in \cite{Hummer:2010vx}, based on the physics of SOPHIA~\citep{Mucke:1999yb}; see also \cite{Murase:2005hy} for their impact. We also include flavor mixing~\citep{Fogli:2012ua}, magnetic field effects on the secondaries, see \cite{Kashti:2005qa,Lipari:2007su,Baerwald:2010fk}, the kinematics of the weak decays~\citep{Lipari:2007su}, and the recent revisions of the normalization in \cite{Hummer:2011ms}. 
For details on the underlying model, see \cite{Baerwald:2011ee}.

\subsubsection*{Acceleration of protons (zone~I)}

We assume that protons are accelerated by Fermi shock acceleration in the acceleration zone to obtain a power law spectrum  with spectral index $\alpha_p \simeq 2$. 
The acceleration rate is empirically described, as usual, by (see, \eg, \cite{Hillas:1985is})
\begin{equation}
 t'^{-1}_{\mathrm{acc},I} = \eta_I \frac{c}{R_L'} = \eta_I \frac{c^2 e B'}{E'} = 9 \cdot 10^3 \, \eta_I \, \frac{B' \, [\text{G}]}{E' \, [\text{GeV}]} 
\label{equ:accI}
\end{equation}
with the acceleration efficiency $0.1 \lesssim \eta_I \lesssim 1$ in that definition. Here, the  energy gain is $t'^{-1} \equiv E'^{-1} |dE'/dt'|$, $\eta_I$ corresponds to the fractional energy gain per cycle, and $R_L'$ to the cycle time. 

The acceleration can only be efficient up to the maximal (or critical) energy, where escape or energy losses start to dominate over the acceleration efficiency. We neglect photohadronic losses since the considered bursts are optically thin to neutron escape. We assume that synchrotron or adiabatic losses ($t'^{-1}_{\text{ad}} \simeq c/\Delta r'$) limit the maximal energy, whatever loss rate is larger, or the dynamical timescale.
We obtain the maximal particle energies from $t^{-1}_{\mathrm{acc}}=t^{-1}_{\mathrm{loss}}$ as 
\begin{eqnarray}
&& E'_{c} \, [\text{GeV}] = \nonumber\\
 &&\min \left( 2.3 \cdot 10^{11} \, (m \, [\text{GeV}])^2 \sqrt{\frac{\eta_I}{B' \, [\text{G}]}} \, ,\right.\nonumber\\
&&  \, \left.9 \cdot 10^3 \eta_I \, B' \, [\text{G}] \, \frac{\Gamma \, t_v \, [\text{s}]}{1+z}  \right)\,.
\label{equ:ec}
\end{eqnarray}
The first entry corresponds to the synchrotron-limited case, and the second entry to the adiabatic loss or dynamical timescale-limited case. Note that \equ{ec} can be applied to the protons as well as the secondaries, at least if they are accelerated by shock acceleration. 

The magnetic field can be estimated by energy partition arguments from the observables as~\cite{Baerwald:2011ee}
\begin{equation}
	B' \simeq 130 \, \left( \frac{\epsilon_B}{\epsilon_e} \right)^{\frac{1}{2}} \, \left( \frac{L_{\text{iso}}}{10^{52} \, \mathrm{erg \, s^{-1}}}\right)^{\frac{1}{2}} \, \left( \frac{\Gamma}{10^{2.5}} \right)^{-3} \, \left( \frac{t_v}{0.01 \, \second} \right)^{-1}  \, \left( \frac{1+\textit{z}}{3} \right) \,  \kilo\text{G}  \, ,
	\label{equ:BISM}
\end{equation}
where $\epsilon_B/\epsilon_e \simeq 1$ describes equipartition between magnetic field energy and kinetic energy of the electrons.
Note that the variability timescale $t_v$ is given in the observer's frame at Earth, not in the source (engine) frame. 

\subsubsection*{Radiation processes and stochastic acceleration of secondaries (zone~II)}

The protons are injected from the acceleration into the radiation zone with a spectrum $Q'_{II,p} \propto (E'_p)^{-\alpha_p}$, where $Q'$ carries units of $[\mathrm{GeV^{-1} \, cm^{-3} \, s^{-1}}]$. In that zone, the protons may interact with photons to produce %SK the 
secondaries, which undergo synchrotron losses, decay, escape (over the dynamical timescale), and adiabatic losses, see \cite{Winter:2012xq} for details. In addition, we consider stochastic acceleration (second order Fermi acceleration) in the spirit of \cite{Murase:2011cx}, following \cite{Weidinger:2010gp,Weidinger:2010px,Murase:2011cx}. If there is substantial turbulence, this stochastic (Fermi 2nd order) acceleration may be important. For the sake of simplicity, we assume that the turbulent region covers the whole zone~II; if only a fraction is affected, only a fraction of particles will be accelerated and one can trivially obtain the results from our figures.

The steady state kinetic equation for the secondary muons, pions, and kaons in zone~II is given by
\begin{equation}
 Q'_{II,i} = \frac{N'_{II,i}}{t'_{\mathrm{esc}}} - \frac{\partial}{\partial E'} \left( \frac{E' N'_{II,i}}{t'_{\mathrm{loss}}} \right) + \frac{\partial}{\partial E'} \left( \frac{E' N'_{II,i}}{t'_{\mathrm{acc},II}} \right) -  \frac{\partial}{\partial E'} \left( \frac{E'^2}{2 t'_{\mathrm{acc},I}} \frac{\partial N'_{II,i}}{\partial E'} \right)\, ,
\label{equ:fermiII}
\end{equation}
where $Q_i'$ the injection of species $i$ from photohadronic processes or parent decays and $N_i'$ is the steady state density (units $[\mathrm{GeV^{-1} \, cm^{-3}}]$). The first term on the r.h.s. describes escape by decay or escape over the dynamical timescale, \ie, $t'^{-1}_{\mathrm{esc}} = t'^{-1}_{\mathrm{decay}}+t'^{-1}_{\mathrm{dyn}}$.
The second term describes energy losses, \ie, $t'^{-1}_{\mathrm{loss}} = t'^{-1}_{\mathrm{synchr}}+t'^{-1}_{\mathrm{ad}}$.  Without acceleration, decay typically dominates at low energies and synchrotron losses at high energies, and for $t'^{-1}_{\mathrm{synchr}} \simeq t'^{-1}_{\mathrm{decay}}$, a spectral break by two powers is expected. The last two terms in \equ{fermiII} are characteristic for stochastic acceleration and always come together with a fixed relative magnitude, see \eg\ \cite{Weidinger:2010gp,Weidinger:2010px}. 

We assume that the acceleration timescale $t'_{\mathrm{acc},II}$ is given by 
 \begin{equation}
 t'_{\mathrm{acc},II} \equiv \frac{E'^2}{2 D'_{EE}} = \tilde \eta_{II}^{-1} \frac{l'_{\mathrm{tur}}}{c} \left( \frac{R'_L}{l'_{\mathrm{tur}}} \right)^{2-q} \simeq \eta_{II}^{-1} t'_{\mathrm{dyn}}  = \eta_{II}^{-1} \, \Gamma  \, \frac{t_v}{1+z} \, ,
\label{equ:accii}
 \end{equation}
following \cite{Murase:2011cx}; see discussion therein.
 The energy diffusion coefficient is assumed to be $D'_{EE} \propto E'^q$, and $l'_{\mathrm{tur}}$ is the length scale of the turbulence -- which can be estimated from the typical lifetime of the turbulence. In the third step, we have chosen $q \simeq 2$~\citep{Murase:2011cx}, and we have re-parametrized the acceleration timescale in terms of the shell width and turbulence length scale as $ \eta_{II} = \tilde \eta_{II} \Delta R' / l'_{\mathrm{tur}}$. 

We expect significant effects of stochastic acceleration if $\eta_{II} > 1$, since then the acceleration exceeds the escape in a certain energy window. Let's consider the most extreme case, the kaons, which have the highest energies at their cooling break.  If these 
are to be accelerated and confined, the condition $R'_L < c  t'_{\mathrm{acc},II} \lesssim l'_{\mathrm{tur}} < \Delta R' \simeq c t'_{\mathrm{dyn}}$ implies that the stochastic acceleration timescale is longer than the shock acceleration timescale but shorter than the hydrodynamical timescale. One finds $1 < \eta_{II} \lesssim 10$ as a reasonable parameter range; \cf, \equ{accii}.\footnote{In the most extreme case, if $\eta_{I}=1$ and the maximal energy is dominated by adiabatic losses, the kaons will take about 35\% of the maximal proton energy. Thus, their Larmor radius will be of the order of one tenth of the size of the region. In other cases and for other species, it will be smaller.} We neglect acceleration of the primary protons in zone~II, since one can show analytically that the Fermi 2nd order acceleration only changes the overall normalization of a simple power law. The proton spectrum normalization is however determined by energy partition arguments, 
so the impact of Fermi 2nd order re-acceleration can be absorbed in a re-definition of the baryonic loading.
 In addition, 
%SK note that our 
the target photon spectrum is based on observation, which means that we do not need to consider the acceleration of electrons in the spirit of \cite{Murase:2011cx} to describe 
%SK
the prompt emission spectrum. As in \cite{Murase:2011cx}, we assume that the secondaries in the relevant energy range cannot escape, since $R'_L < \Delta R'$.

\subsubsection*{Transport of secondaries back to acceleration zone~I}

Apart from stochastic acceleration in the radiation zone, it is conceivable that a substantial fraction of secondaries is transported back to the acceleration zone~I by diffusion. We characterize this fraction as $f_{\text{diff}} \simeq \lambda'/\Delta R'$, where $\lambda'$ is the diffusion length over the dynamical timescale. Our description closely follows \cite{Baerwald:2013pu} in that aspect, and we 
 assume that the secondaries are produced uniformly over the radiation zone~II. 

The fraction $f_{\text{diff}}$ of particles which can diffuse back to the shock front within the dynamical timescale can be estimated from the diffusion length $\lambda' \simeq \sqrt{D'_{xx} \, t'_{\mathrm{dyn}}}$ as
\begin{equation}
 f_{\text{diff}} = \min \left( \frac{\lambda'}{c t'_{\mathrm{dyn}}} , 1 \right) \, ,
\label{equ:fdiff}
\end{equation}
where $D'_{xx}$ is the spatial diffusion coefficient. This definition ensures that  $f_{\text{diff}} \le 1$.
For example, for Bohm-like diffusion, one has $D'_{xx} \propto E'$ and for Kolmogorov-like diffusion, one has $D'_{xx} \propto E'^{1/3}$ \citep{stanev2003,schlickeiser2002}, and as a consequence, $f_{\text{diff}} \propto \sqrt{E'}$ and $f_{\text{diff}} \propto E'^{1/6}$, respectively. As a lower limit, it can be shown that a fraction $f_{\text{dir}}=R'_L/(c t'_{\mathrm{dyn}}) \propto E'$ of the secondaries can directly escape from the radiation zone in the same way as cosmic rays from the shells~\citep{Baerwald:2013pu}. That is, when the Larmor radius becomes comparable to the shell width, all particles will reach back to the shocks. It is therefore reasonable to normalize the transport back to zone~I in the way that for $R'_L=\Delta R'$ all particles are efficiently transported.\footnote{That is, we choose $f_{\text{diff}}=(R'_L/(c t'_{\mathrm{dyn}}))^\gamma$ with $D'_{xx} \propto E'^{2 \gamma}$.  } 

Muons, pions, and kaons will typically not reach these high energies, since synchrotron losses lead to a spectral break. As a consequence, only the fraction $f_{\text{diff}} \simeq \sqrt{E'_{\mathrm{break}}/E'_{\mathrm{c}}}$ will diffuse back, at the most, where $E'_{\mathrm{c}}$ is the maximal energy in \equ{ec}. 
For pions and kaons the break energies are typically higher than for muons, which means that %SK we expect 
a larger fraction of pions and kaons should be transported back to zone~I.

Note that  stochastic acceleration and the transport by diffusion are connected via transport theory. Specifically, it is shown in e.g.\ \cite{schlickeiser2002}, that in the relativistic limit of $E\approx p\cdot c$, the product of the spatial and momentum diffusion coefficients is given as 
\begin{equation}
D'_{EE}\cdot D'_{xx}=\frac{4E^2\,v_{A}^{2}}{3\,a\,(4-a^2)\,(4-a)\,w}\,.
\end{equation}
where $w$ is a constant parameter defining the turbulence scale, which
is often included in the definition of the Alfv{\'e}n velocity $v_{A}$
(see e.g.\ \cite{gebauer_phd2010} for a summary). 
The wave spectrum follows a power law $k^{a}$ with the index
$a$ connected to the spatial diffusion
coefficient as $D_{xx}\propto E'^{2-a}$. For Kolmogorov-type
diffusion, $a=5/3$, while in the Bohm-case, $a=1$. This means that
efficient stochastic acceleration $t'_{\mathrm{acc},II} \propto {D'_{EE}}^{-1}$ in zone~II, see \equ{accii},
 implies inefficient spatial transport, and vice versa. In particular, if $D'_{EE} \propto {E'}^{q}$, as we assumed above, $D'_{xx} \propto {E'}^{2-q}$. Therefore, $q \sim 2$ is roughly consistent with Kolmogorov diffusion, which we use as a standard in the following. It is conceivable from this discussion that the acceleration of the secondaries dominates either in zone~I or zone~II as a function of energy, depending on the efficiency of transport back to the acceleration zone versus stochastic acceleration.

\subsubsection*{Shock acceleration of secondaries in zone~I}

The injection from the radiation back into the shock zone is given by
\begin{equation}
 Q'_{I,i}=N'_{II,i} \, t'^{-1}_{\mathrm{dyn}} \, f_{\text{diff}}(E')=N'_{II,i} \, t'^{-1}_{\mathrm{eff,diff}}  \, ,
\label{equ:inj}
\end{equation}
where one can define the effective diffusion timescale $t'^{-1}_{\mathrm{eff,diff}} \equiv  t'^{-1}_{\mathrm{dyn}} f_{\text{diff}}$. Note that $N'_{II,i} t'^{-1}_{\mathrm{dyn}}$ is the ejected spectrum if all particles can escape from zone~II over the dynamical timescale, whereas $f_{\text{diff}} \le 1$ characterizes the energy-dependent fraction obtained from \equ{fdiff}. In addition, note that $t'^{-1}_{\mathrm{eff,diff}} \le  t'^{-1}_{\mathrm{dyn}} \simeq t'^{-1}_{\mathrm{ad}}$, 
so diffusion is always less efficient than the adiabatic cooling or escape over the dynamical timescale in zone~II, and is therefore not included in \equ{fermiII}.

The corresponding kinetic equation for the secondaries in zone~I is given in the steady state by
\begin{equation}
 Q'_{I,i} = \frac{N'_{I,i}}{t'_{\mathrm{esc}}} - \frac{\partial}{\partial E'} \left( \frac{E' N'_{I,i}}{t'_{\mathrm{synchr}}} \right) + \frac{\partial}{\partial E'} \left( \frac{E' N'_{I,i}}{t'_{\mathrm{acc},I}} \right) \, 
\end{equation}
with the same acceleration efficiency as for the protons \equ{accI}. 
That is, we assume that the secondaries undergo acceleration similar to the protons, suffer from synchrotron losses, and escape via decay and escape from the acceleration zone over the dynamical timescale, \ie, $t'^{-1}_{\mathrm{esc}} = t'^{-1}_{\mathrm{dyn}}+t'^{-1}_{\mathrm{decay}}$.  Since  $t^{-1}_{\mathrm{acc},I} > t^{-1}_{\mathrm{decay}}$ in order to have significant secondary acceleration (both have the same energy dependence), the particles at the highest energies can typically escape over the dynamical timescale from zone~I before they decay. Therefore, we assume that accelerated secondaries decay in the radiation zone.\footnote{That is only relevant for accelerated pions which may decay in the acceleration zone, such that the resulting muons are guaranteed to be in the shock from the beginning.}

One may ask if this approach is consistent with the textbook version of Fermi shock acceleration. In that version, the proton index is given by $\alpha_p = P_{\mathrm{esc}}/\eta_I+1$, where $P_{\mathrm{esc}}$ is the (constant) escape probability per cycle and $\eta$ is the (constant) fractional energy gain per cycle. The ratio $P_{\mathrm{esc}}/\eta_I = 3/(\chi-1) \simeq 1$ depends on the compression ratio $\chi$ only, where $\chi \simeq 4$ for a strong shock. As a consequence, a ``intrinsic'' escape term $t^{-1}_{\mathrm{esc,shock}}=t^{-1}_{\mathrm{acc}}$ is needed for a self-consistent kinetic simulation. In our approach, we checked analytically and numerically that such an additional escape term $t'^{-1}_{\mathrm{esc,shock}} \simeq  P_{\mathrm{esc}}/T'_{\mathrm{cycle}} \propto E'^{-1}$ with $P_{\mathrm{esc}}=\eta_I$ and $T'_{\mathrm{cycle}} \simeq R_L'/c$ produces an $E'^{-2}$ ejection spectrum for the protons if a narrow-energetic particle distribution is injected. Here,  it is crucial that acceleration and escape terms carry the same energy dependence (which is implied by the constant energy gain and escape probability per cycle), and that $Q'_{\mathrm{esc}}=N'/t'_{\mathrm{esc,shock}}$, which means that $Q'_{\mathrm{esc}}$ and $N'$ have different energy dependencies. For the secondaries, such an escape term will suppress the spectra somewhat, depending on the spectral index of the injection (determined by the ratio $P_{\mathrm{esc}}/\eta_I$).
For the sake of simplicity, we  assume that the secondaries will escape via decay or over the dynamical timescale only. This is in a way the most aggressive assumption one can make, which will however support our conclusions. It may also apply if the escape properties change over time, the acceleration site of the secondaries is different from the one of the primaries, or if the secondaries, which have lower energies than the protons, are trapped in magnetic fields, whereas the protons are injected into the shock at relatively high energies with a larger Larmor radius. 

\section{Impact of acceleration effects on the secondaries}

\begin{table}
\caption{\label{tab:TblA} Properties of four bursts discussed in this study, see \cite{Baerwald:2010fk} for SB (``Standard Burst'', similar shape to \cite{Waxman:1997ti,Waxman:1998yy}), \cite{Nava:2010ig,Greiner:2009pm} for GRB080916C, \cite{Nava:2010ig,Abdo:2009pg} for GRB090902B and \cite{Nava:2010ig,Gruber:2011gu} for GRB091024. The luminosity is calculated with $L_\text{iso}=4\pi d_L^2\cdot S_\gamma/T_{90}$, with $S_\gamma$ the fluence in the (bolometrically adjusted) energy range $1\,\text{keV}-10\,\text{MeV}$. For the gamma-ray spectrum, a broken power law is assumed with spectral index $\alpha_\gamma$ below the break, $\beta_\gamma$ above the break, and the break energy $\epsilon_{\gamma,\text{break}}$. Adopted from \cite{PhDHummer}.}             % title of Table
\centering                          % used for centering table
\begin{tabular}{c|c c c c}        % centered columns (5 columns)
\hline\hline                 % inserts double horizontal lines
                           GRB              & SB                & 080916C          & 090902B          & 091024           \\
\hline                        % inserts single horizontal line
  $\alpha_\gamma$                        & 1                 & 0.91                & 0.61                & 1.01                \\
  $\beta_\gamma$                         & 2                 & 2.08                & 3.80                & 2.17                \\
  $\epsilon_{\gamma,\text{break}}$ [MeV] & 1.556             & 0.167               & 0.613               & 0.081               \\
  $\Gamma$                               & $10^{2.5}$        & 1090                & 1000                & 195                 \\
  $t_v$ [s]                              & 0.0045            & 0.1                 & 0.053               & 0.032               \\
  $T_{90}$ [s]                           & 30                & 66                  & 22                  & 196                 \\
  $z$                                    & 2                 & 4.35                & 1.822               & 1.09                \\
  $S_\gamma$ [erg cm$^{-2}$]   & $1 \cdot 10^{-5}$ & $1.6 \cdot 10^{-4}$ & $3.3 \cdot 10^{-4}$ & $5.1 \cdot 10^{-5}$ \\
  $L_\text{iso}$ [erg s$^{-1}$]   & $10^{52}$         & $4.9 \cdot 10^{53}$ & $3.6 \cdot 10^{53}$ & $1.7 \cdot 10^{51}$ \\
\hline                                   %inserts single line
\end{tabular}
\end{table}

\begin{figure*}
\centering
\includegraphics[width=16.4cm,clip]{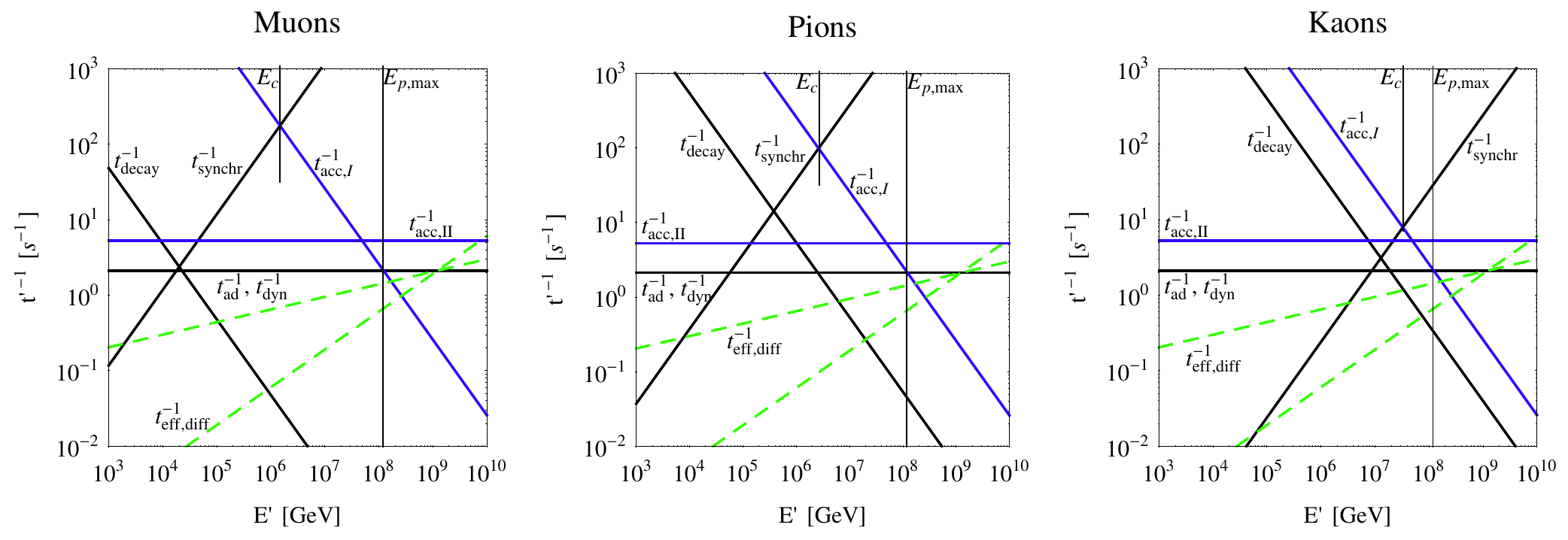}
\caption{Relevant inverse timescales for secondary muons, pions, and kaons (in the different panels) as a function of the secondary energy in the SRF. The chosen acceleration rate for zones~I and~II are $\eta_I=0.1$ and $\eta_{II}=2.5$, respectively. The burst parameters correspond to the Standard Burst in \Tab~\ref{tab:TblA}.}
\label{fig:timescales}
\end{figure*}

In order to illustrate the impact of the acceleration on the secondaries, we choose the GRB parameters listed 
in the second column of \Tab~\ref{tab:TblA} for a burst  chosen to reproduce the properties of the Waxman-Bahcall burst~\citep{Waxman:1997ti,Waxman:1998yy} (plateau between $10^5$ and $10^7 \, \mathrm{GeV}$ in $E_\nu^2 F_\nu$), see \cite{Baerwald:2010fk}. The corresponding inverse timescales are shown in \figu{timescales} for the secondary muons, pions, and kaons (in the different panels, in SRF). We can use this figure to discuss the expected behavior in the different zones.

In the acceleration zone (I), acceleration or synchrotron losses dominate for all species. The maximal (critical) energy can be obtained from \equ{ec} as for protons, where adiabatic losses are also included as possibility to limit the maximal energy.
From the figure it is clear that it is close to each other for muons and pions, whereas it is significantly higher for kaons.   Here, all secondary species can be, in principle, efficiently accelerated in the shock, since  $t'^{-1}_{\mathrm{decay}} \ll t'^{-1}_{\mathrm{acc},I}$. The largest difference between decay and acceleration, which have the same energy dependence, is obtained for muons, the smallest for kaons. Therefore, one may expect that muons are most efficiently accelerated, see also \cite{Klein:2012ug}. 

The pile-up depends on the energy efficiency range of the acceleration. For GRBs, that is non-trivial to determine, since the secondary spectrum has a spectral break coming from the the gamma-ray spectrum; hence the potential pile-up range is given by the interval between that break and the critical energy. Another break, the synchrotron cooling break, can be obtained from $t'^{-1}_{\mathrm{decay}}=t'^{-1}_{\mathrm{synchr}}$, and is lowest for muons and highest for kaons. 
It shows up in all cases at lower energies than $E_c$. 
Even more complicated, the critical energy is above the cooling break in all cases, which means that its energy is beyond the peak energy of the spectrum, and that it is not guaranteed that the peak flux of the spectrum will be increased at the absolute maximum.  Note that it is not simply possible to lower the magnetic field to reduce the cooling and enhance the effect of the acceleration, since the acceleration efficiency will be reduced, whereas the cooling break will persist as adiabatic cooling break even if synchrotron losses are suppressed (where $t'^{-1}_{\mathrm{decay}}=t'^{-1}_{\mathrm{ad}}$). We will however discuss the conditions for the possibly largest acceleration effects in the next section. 

As far as the transport between radiation zone, where the secondaries are mostly produced, and the acceleration zone is concerned, we show the effective diffusion rates $t'^{-1}_{\mathrm{eff,diff}}$ (\cf, \equ{inj}) for the Kolmogorov and Bohm cases in the figure as upper and lower dashed curves, respectively. It is clear that the higher the critical energy, the more particles will be transported back to the shock. Therefore, the transport is expected to be most efficient for kaons, which somewhat compensates for the less efficient acceleration -- depending on the transport type. The Kolmogorov and Bohm cases give the range of plausible transport scenarios. Perfect transport (all particles transported back to the shock over the dynamical timescale) would correspond to  $t'^{-1}_{\mathrm{eff,diff}}=t'^{-1}_{\mathrm{dyn}}$. As most conservative assumption, only the particles not scattering at all may reach back to the shock, corresponding to the direct escape in \cite{Baerwald:2013pu}. In that case, the transport is only efficient if the Larmor radius reaches the size of the region. We checked that the results for the Kolmogorov case are already quite similar to the perfect transport case, whereas the Bohm case and steeper energy dependencies lead to very small amounts of secondary acceleration, see below.

For the stochastic acceleration in zone~II, the largest effects are expected if $t^{-1}_{\mathrm{acc},II}$ dominates over the synchrotron and decay timescales in the radiation zone. Because of the shallow dependence on energy, a small window (about one order of magnitude in energy) can be found for muons and kaons in \figu{timescales}, whereas pions are hardly affected for the chosen acceleration efficiency. In summary, we expect the most interesting results for muons, which may be efficiently accelerated in both zones, and kaons, which may be efficiently transported back to the shock and efficiently accelerated in the radiation zone.
\begin{figure*}
\centering
\includegraphics[width=0.4\textwidth]{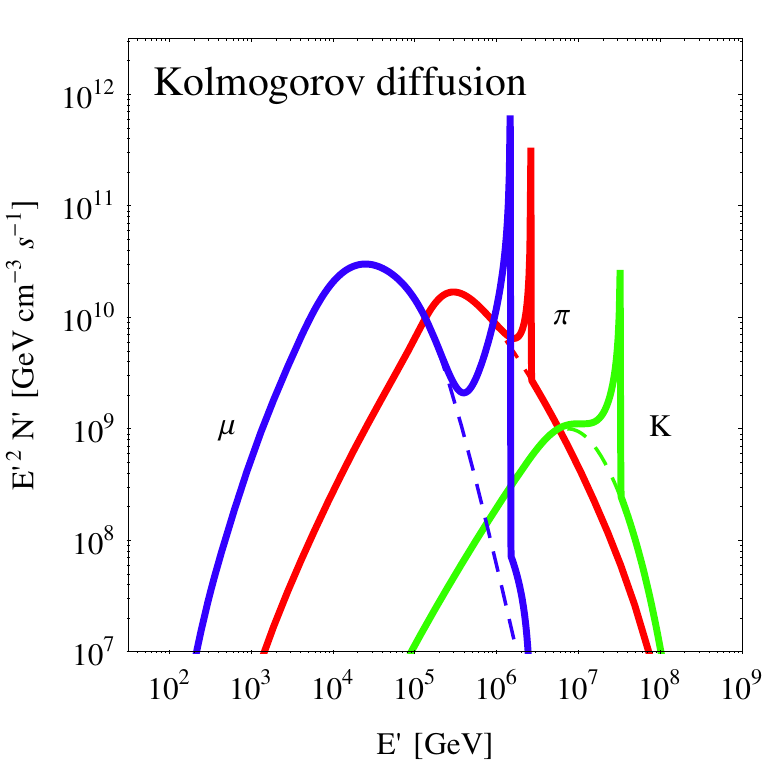}  \hspace*{1cm} %
\includegraphics[width=0.4\textwidth]{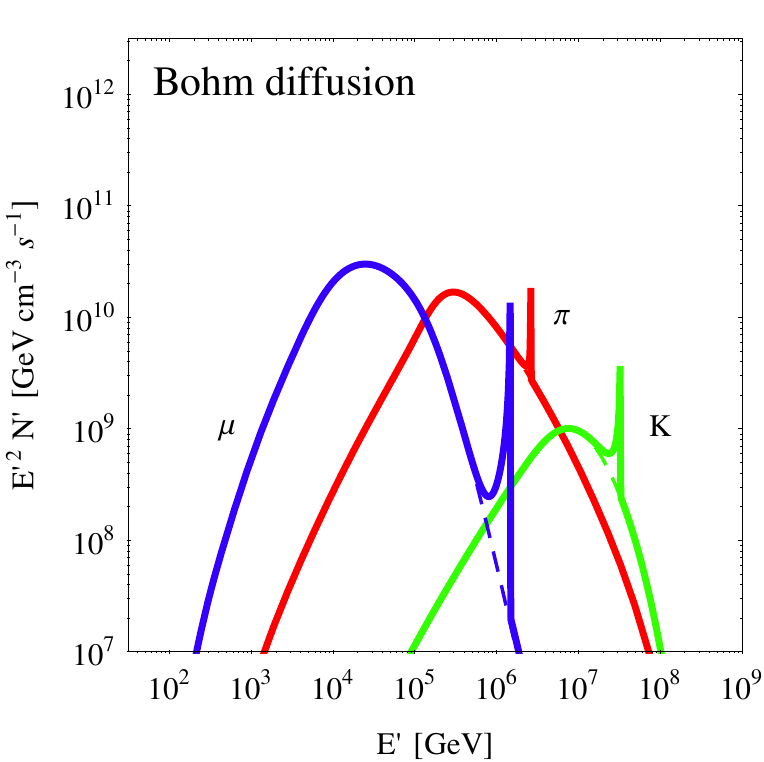}  
\caption{\label{fig:intspectra}  Effect of shock acceleration on steady state densities $N'$ of muons, pions, and kaons for two different transport mechanisms between radiation and acceleration zone (different panels, in SRF). The plots use the burst SB from \Tab~\ref{tab:TblA} and $\eta_I=0.1$. The dashed curves are shown without acceleration for comparison.}
\end{figure*}

These qualitative considerations are quantitatively supported by numerical simulations. Let us focus on \figu{intspectra} first, where the effect of shock acceleration only is shown on the secondary muons, pions, and kaons. 
This figure shows the steady state spectra $N'$, which differ in shape
from the neutrino ejection spectra; see also App.~A in \cite{Baerwald:2011ee}. Solving \equ{fermiII} for decay only, one obtains $Q'=(t'_{\mathrm{decay}})^{-1} N'$ below the peak of the spectra. Therefore, $Q' \propto (E')^{-2}$ for $N' \propto (E')^{-1}$, which can be compared to the neutrino ejection spectra in terms of shape.

The left panel in \figu{intspectra} uses Kolmogorov diffusion from the radiation to the acceleration zone, the right panel Bohm diffusion, where these may be regarded as the optimistic and conservative cases for the transport. The shock acceleration leads to the pile-up spikes at the critical energies, marked in \figu{timescales}, which are also observed in \cite{Klein:2012ug}. The acceleration components are most prominent for Kolmogorov diffusion, where
many secondaries are transported back to the shock, and least prominent for the Bohm diffusion. As we discussed above, because of the balance between transport and acceleration efficiency, muons and kaons are mostly accelerated, whereas the effect on the pions is smaller.

In either case, the spikes in the muon or pion spectra are washed out in the neutrino spectrum, because the kinematics of the weak decays re-distributes the parents' energies.
The spikes lead to shoulders in the neutrino spectra, as can be seen in \figu{allbursts}.
Most importantly, the effect of muon acceleration may be shadowed by the regular pion spectrum, as it is evident from the right panel. Therefore, in the Bohm case, the effect of acceleration on the neutrinos is hardly visible. In the Kolmogorov case, on the other hand, two distinctive peaks (from pion/muon acceleration and from kaon acceleration) should be visible, where the one from pion/muon acceleration is closest to the overall peak of the spectrum and therefore perhaps easiest to detect. In the following, we will only discuss the transport by Kolmogorov diffusion, which may be optimistic but is the minimal requirement to observe significant effects on the neutrino spectra. As a minor detail, 
in \figu{intspectra}, left panel, a small enhancement above the critical energy for muons, which comes from the fact that 
some of the muons are injected above the critical energy from accelerated pions.

\begin{figure*}[tp]
\centering
\includegraphics[width=\textwidth]{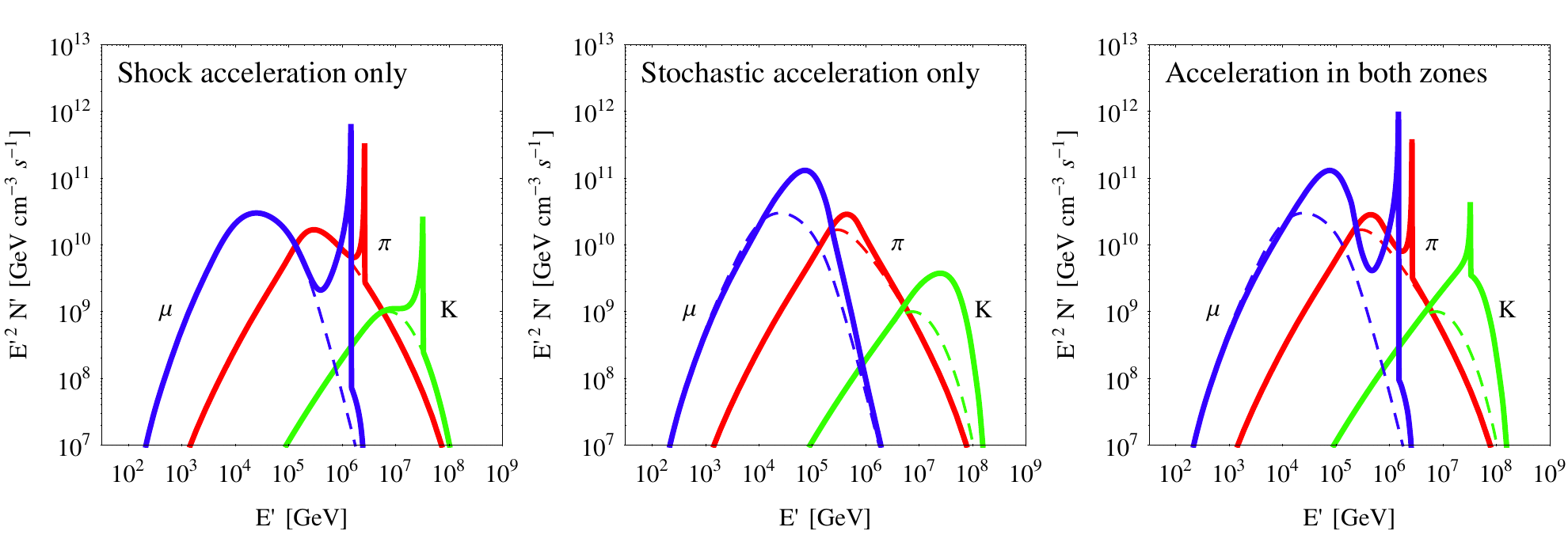} 
\caption{\label{fig:intspectrac} Effect of shock acceleration, stochastic acceleration, and both accelerations combined (in different panels) on steady state densities $N'$ for muons, pions, and kaons. Here, the burst SB from \Tab~\ref{tab:TblA}, $\eta_I=0.1$, and $\eta_{II}=2.5$ have been used. The different panels are for acceleration in the shock (left), in the radiation zone (middle), and both (right). Here, Kolmogorov diffusion has been used as transport mechanism between the two zones. 
The dashed curves are shown without acceleration for comparison.}
\end{figure*}

Apart from shock acceleration of secondaries transported back to the shock, stochastic acceleration in a turbulent radiation zone could be relevant. In order to maximize the effect, we assume that the whole radiation zone is turbulent, and show the impact of shock acceleration (left panel), stochastic acceleration (middle panel), and acceleration in both zones (right panel) in \figu{intspectrac}. As discussed above, stochastic acceleration can lead to a significant enhancement of the muon and kaon spectra at their peaks, where stochastic acceleration is efficient over about one order of magnitude in energy for the chosen acceleration efficiency. The combined effect of acceleration in both zones is shown in the right panel, and is (to a first approximation) an addition of the two effects. 
One could in principle assume even somewhat more extreme acceleration efficiencies in zone~II, which leads to a much stronger enhancement. 
However, current neutrino data~\citep{Abbasi:2012zw} already puts constraints on scenarios with more optimistic secondary acceleration.

\section{Impact on neutrino fluences}

\begin{figure*}[tp]
\centering
\includegraphics[width=0.8\textwidth]{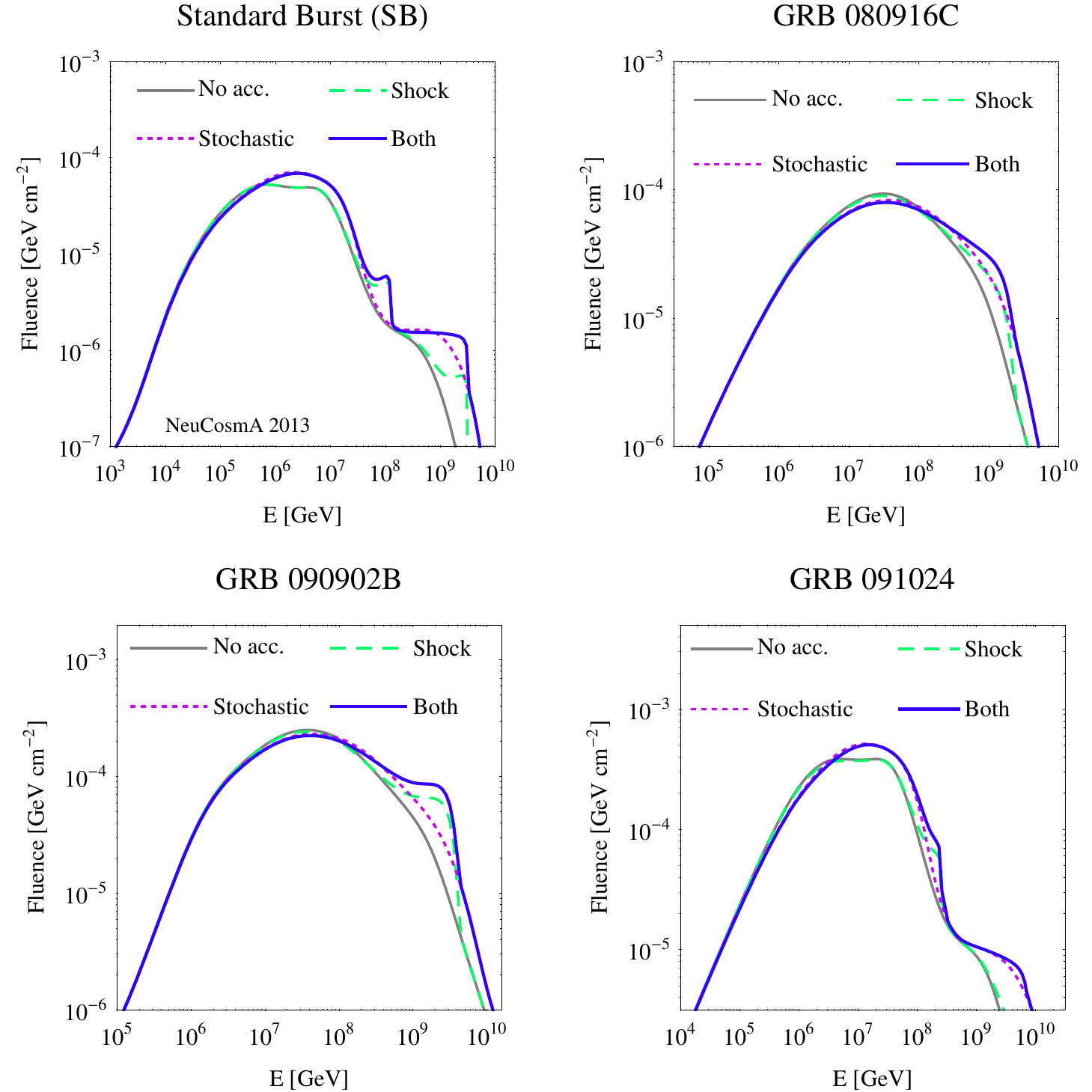} 
 \caption{\label{fig:allbursts} Acceleration on the muon neutrino fluences in the observer's frame for the bursts in \Tab~\ref{tab:TblA}.
Here, $\eta_I = 0.1$ and $\eta_{II}=2.5$ have been assumed, as well as Kolmogorov diffusion as transport mechanism between zones~II and~I. Here, ``No acc.'' refers to no acceleration of the secondaries, ``Shock'' to shock acceleration in zone~I only, ``Stochastic'' to stochastic acceleration in zone~II only, and ``Both'' to the combined effect in both zones. The flavor mixing has been computed with the parameters in \cite{Fogli:2012ua}.
}
\end{figure*}

The upper left panel of \figu{allbursts} shows the neutrino spectra for the standard burst from \Tab~\ref{tab:TblA}. In the muon neutrino fluence, the enhancement of the peaks from muon and kaon decays in the case of stochastic acceleration can be clearly seen. For the shock acceleration, the spikes in \figu{intspectrac} translate into peaks at energies higher by a factor of $\Gamma/(1+z)$. The combined effect enhances the neutrino spectrum by about 50\% in that case, 
leading to an additional peak at about $10^8 \, \mathrm{GeV}$, and increases the neutrino peak from kaon decays significantly. 
The spiky secondary particle spectra lead to  
$E^{-1}$ spectra for the neutrinos, since the kinematics of weak decays cannot exceed this spectral index.

It is, of course, an interesting question how much these observations
depend on the parameter values. We therefore choose three different,
recently observed (by Fermi) GRBs as examples: GRB 080916C, GRB
090902B, and GRB 091024. GRB 080916C is one of the brightest bursts
ever seen, although at a large redshift, and one of the best studied Fermi-LAT bursts. The gamma-ray spectrum of GRB 090902B has a relatively steep cutoff, and might therefore be representative for a class of bursts for which the gamma-ray spectrum can be fit with a single power law with exponential cutoff as well. GRB 091024 can be regarded as a typical example representative for many Fermi-GBM bursts~\citep{Nava:2010ig}, except for the long duration. Note that GRB 080916C and  GRB 090902B have an exceptionally large $\Gamma \gtrsim 1000$, whereas $\Gamma \simeq 200$ for the last burst. All three observed bursts have in common that the required parameters for the neutrino flux computation can be taken from the literature; see the \Tab~\ref{tab:TblA} and its caption for the references. Note that these bursts have been also studied in the context of neutrino decays~\citep{Baerwald:2012kc} and the normalization question~\citep{PhDHummer,Winter:2012xq}.

We show in \figu{allbursts} the neutrino spectra for the four representative GRBs listed in \Tab~\ref{tab:TblA}, 
The effects of the secondary acceleration on the neutrino spectra are depicted in \figu{allbursts}. To a first approximation, the effects are 
dominated by the strength of the magnetic field strength, which can be estimated with \equ{BISM}. GRB 091024 has a similar magnetic field (about 60~kG) to our Standard Burst (about 290~kG), whereas the magnetic fields for GRB 080916C (4~kG) and GRB 090902B (6~kG) are significantly lower because of their large Lorentz boosts. Consequently, the spectral shapes of the neutrino spectra are very different, dominated by an adiabatic cooling break which changes the spectrum only by one power. In these cases, it is possible that the stochastic and shock acceleration effects add up.

One may ask the question when these largest effects can be expected and if they can be enhanced. The peak of the secondary spectrum in $E^2 N'$ is given by the cooling break $t'^{-1}_{\mathrm{synchr}}=t'^{-1}_{\mathrm{decay}}$. The critical energy for the shock acceleration is typically given by $t'^{-1}_{\mathrm{synchr}}=t'^{-1}_{\mathrm{acc},I}$. The critical energy for the stochastic acceleration is determined by  $t'^{-1}_{\mathrm{synchr}}=t'^{-1}_{\mathrm{acc},II}$, which means that stochastic acceleration can be efficient up to relatively large energies for small $B'$ (as one can see in the figure). One 
expects the maximal enhancement effect at the peak for  $t'^{-1}_{\mathrm{decay}}=t'^{-1}_{\mathrm{acc,I}} \simeq t'^{-1}_{\mathrm{synchr}}$, where the cooling break and critical energy for shock acceleration coincide. This condition translates into a critical magnetic field
\begin{equation}
B_c' \simeq 10^{-4} \, \eta_I^{-1} \frac{m \, [\mathrm{GeV}]}{\tau_0 \, [\mathrm{s}]} \, \mathrm{G} \, ,
\end{equation}
where $m$ is the mass of the secondary and $\tau_0$ its rest-frame lifetime. The 
ratio $m/\tau_0 \simeq 4.8 \cdot 10^{4} \, \mathrm{GeV \, s^{-1}}$ is smallest for muons, where $B_c' \simeq 50 \, \mathrm{G}$ (for $\eta_I=0.1$). Since for muons, the acceleration is most efficient, the pions and kaons will rather decay than being accelerated in that case.  The closest parameters can be found for the GRBs with high Lorentz factors (to achieve high energies) and relatively low magnetic fields GRB080916C and GRB090902B, best seen in the lower left panel of \figu{allbursts} as additional peak only a factor of few above the energy of the absolute maximum.
In principle, one can also find such a critical magnetic field for kaons, for which 
$m/\tau_0 \simeq 4 \cdot 10^{7} \, \mathrm{GeV \, s^{-1}}$ 
so $B_c' \simeq 40 \, \mathrm{kG}$ (for $\eta_I=0.1$). This case is not so far away from 
what is shown in \figu{intspectra} ($B' \simeq 290 \, \mathrm{kG}$). However, the kaon peak is far away from the absolute peak of the spectrum. For pions, the spectral peak is closer to that of the muons and the acceleration is less efficient. 
Again, one cannot arbitrary reduce the magnetic field, since low $B'$
mean low proton acceleration efficiencies and low maximal energies. In
the shown cases, there is a balance between a low $B'$ and high
energies obtained by high $\Gamma$, which cannot be separated independently. Finally, there is some impact of the acceleration efficiencies $\eta_I$ and $\eta_{II}$ on the neutrino result, which shift the peaks as qualitatively expected.

\section{Summary and conclusions}

The aim of this study has been to address the quantitative importance
of the acceleration of secondary muons, pions, and kaons for the
neutrino fluxes. We have therefore extended the model by \cite{Hummer:2011ms}, which predicts the neutrino
fluxes from gamma-ray observations in the internal shock model and
which the current state-of-the-art GRB stacking analyses in neutrino
telescopes are based on, by the acceleration effects of the
secondaries -- as discussed in a more general sense in \cite{Klein:2012ug}.

One of the key issues has been a separate description of the acceleration zone (the shocks) and the radiation zone (the plasma downstream the shocks) in a two-zone model, since it is plausible that the shock acceleration and the photohadronic processes, leading to the secondary production, happen dominantly in different regions. Two classes of acceleration have been implemented for the secondaries: shock acceleration in the acceleration zone and stochastic acceleration in the (possibly turbulent) plasma in the radiation zone. An important component of the model has been the transport of the secondaries from the radiation zone back to the acceleration zone, which we describe by Kolmogorov diffusion (optimistic) or Bohm diffusion (conservative) -- assuming that at the highest energies, where the Larmor radius reaches the size of the region, all secondaries are efficiently transported.
The shock acceleration of the secondaries is then just a consequence of the efficient proton acceleration if they can be transported back to the shocks, whereas the stochastic acceleration depends on the size of the turbulent region. In both cases, some uncertainty arises from the acceleration efficiencies, which may vary within reasonable limits.

We have shown that both the muon and kaon spectra can be significantly modified by shock acceleration: the muon spectrum, because muons have a long lifetime over which they can be accelerated, and the kaon spectrum, because kaons are most efficiently transported back to the acceleration zone at their highest energies (they have the highest synchrotron cooling break). The shock acceleration leads to additional peaks determined by the critical energy, where acceleration and energy loss or escape rates are equal. These peaks translate into corresponding peaks of the neutrino spectra, smeared out by the kinematics of the weak decays.

The most significant enhancement at the peak is expected from the muon spectrum if the magnetic field is low and the Lorentz boost is high, since then the critical energy may coincide with the peak energy. Too low magnetic fields, on the other hand, mean that the protons cannot be efficiently accelerated.  We note that our model is fully self-consistent in the sense that it is taken into account that muons are produced by pion decays, which may be accelerated themselves.

The amount
of shock acceleration depends critically on the transport between radiation and acceleration zone. For Bohm diffusion or even slower transport processes, hardly any modification of the neutrino spectra is observed, since the enhancement of the muon spectrum is completely shadowed by the regular pion spectrum present at higher energies. On the other hand, the results for Kolmogorov diffusion are already close to the perfect transport case (all particles efficiently transported over the dynamical timescale). 

The stochastic acceleration can be very efficient for muons and kaons, since their cooling breaks occur at a smaller (decay and synchrotron loss) rates than the one for pions, which means that the stochastic acceleration can be dominant at these breaks. The consequence is an enhancement at the cooling break (if the break comes from synchrotron losses) or beyond (if it comes from adiabatic losses). In the latter case, the shock and stochastic acceleration effects can add up and lead to an additional peak in the neutrino spectrum with a significant enhancement. It is conceivable that efficient stochastic acceleration means inefficient transport, \ie, the two acceleration effects are mutually exclusive in terms of their energy ranges, and that such an effect can be only observed for a flat enough energy dependence of the diffusion coefficient (such as Kolmogorov diffusion).

Depending on the specific GRB parameters, secondary particle acceleration can enhance the neutrino flux by up to an overall factor of two.   The enhancement is typically largest at higher energies, around $10^{8}$ GeV or above. 
This enhancement is relevant for extremely high-energy searches with IceCube at energies around $10^{8}$~GeV and above. In particular, southern hemisphere searches are sensitive at these energies, as the background of atmospheric muons is sufficiently small at those high energies, see \cite{Aartsen:2013ApJ} for the latest point source sensitivity of IceCube. Even northern hemisphere searches might already be sensitive to the enhancement. Next generation instruments like KM3NeT and a high-energy extension of IceCube will be able to constrain the parameter space for secondary particle acceleration effects even further.
Other future experiments which have good sensitivity above $10^{8}$~GeV concern the radio emission from neutrino-induced showers, such as ARA \citep{Allison:2011wk} and ARIANNA \citep{Gerhardt:2010js,Klein:2012bu}.   

These conclusions will somewhat depend on the choices of the acceleration rates, which means that we cannot exclude larger effects for individual bursts. Note that some of our choices (such as for transport and size of the turbulent region) are already on the optimistic side.

%%%%%%%%%%%%%%%%%%%%%%%%%%%%%%%%%%%%%%%%%%%%%%%%%%%%%%%%%%%%5
%%%%%%%%%%%%%%%%%%%%%%%%%%%%%%%%%%%%%%%%%%%%%%%%%%%%%%%%%%%%%%

\begin{acknowledgements}
WW acknowledges support from DFG grants WI 2639/3-1 and WI 2639/4-1, the FP7 Invisibles network (Marie Curie
Actions, PITN-GA-2011-289442), and the ``Helmholtz Alliance for Astroparticle Physics HAP'', 
funded by the Initiative and Networking fund of the Helmholtz
association. JBT acknowledges support from the Research Department of Plasmas with Complex Interactions (Bochum) and from the MERCUR Project Pr-2012-0008.
The work of SK was supported in part by U.S. National Science Foundation under grant 
PHY-1307472 and the U.S. Department of Energy under contract number DE-AC-76SF00098.
We are grateful to P. Baerwald, R.\ Tarkeshian, and E. Waxman 
for useful discussions.  We thank members of the IceCube collaboration for useful discussions. 
\end{acknowledgements}

\end{document}